\documentstyle[twocolumn,aps,pre,graphicx,amssymb,amsbsy,amsfonts]{revtex}

\begin{document}

\title{Criticality of rumor propagation on small-world networks}

\author{Dami\'an H. Zanette}

\address{Consejo Nacional de
Investigaciones Cient\'{\i}ficas y T\'ecnicas\\ Centro At\'omico
Bariloche and Instituto Balseiro, 8400 Bariloche, R\'{\i}o Negro,
Argentina}

\date{\today}
\maketitle

\begin{abstract}
We  show that a simple model for the  propagation  of  a  rumor
on a small-world network exhibits critical  behavior  at a finite
randomness of the underlying graph. The transition occurs between
a regime where the  rumor ``dies'' in a small neighborhood of its
origin, and a regime where  it spreads over a  finite  fraction
of the whole population. Critical exponents are evaluated through
finite-size scaling analysis, and the dependence of  the critical
randomness with the network connectivity is studied. The
behavior  of this system as  a function  of the network
randomness bears noticeable similarities with an epidemiological
model reported recently [M. Kuperman and G. Abramson,  Phys. Rev.
Lett.  {\bf 86}, 2909 (2001)], in spite of substantial
differences in the respective dynamical rules.
\end{abstract}


\vspace{5 pt}

Small-world  networks  have  been  introduced  as an interpolation
between  ordered  and  random  graphs,  to  capture  two  specific
features  of  real  neural,  social,  and  technological  networks
\cite{1,2}.  On  one hand, they have a relatively large clustering
coefficient,  i.e.  the  probability that two neighbors of a given
vertex  are  in  turn  mutual  neighbors  is  high,  as in ordered
graphs.  On the other, the typical separation between two vertices
is  much  smaller  than the total number of vertices, as in random
graphs.   Whereas   the   geometrical  properties  of  small-world
networks  have  been  studied  in detail
\cite{1,2,2a,3,4,5,6,7,8,9}, less attention has been paid to the
dynamical properties resulting from   their   partially  random
structure  \cite{2,9,10,11}.  As explained  in  more  detail
below,  the  structure of small-world networks  is  parametrized
by the randomness $p$, where $p=0$ and $p=1$   correspond   to
fully   ordered   and   random  neworks, respectively.  It  has
been  shown  that, in asymptotically large small-world networks,
statistical  geometrical properties---such as  the mean distance
between vertices---exhibit qualitatively the same behavior  for
any  $p>0$.  Here,  we study an epidemic-like propagation
process  on  a small-world network which, in contrast with
geometrical  properties, shows a transition between two
qualitatively  different  dynamical  regimes  at a finite value of
$p$.  Since  this model addresses a social process and small-world
networks  are  a  presumably  realistic  representation  of social
networks,  the  analysis  could  be relevant to the description of
threshold phenomena in real societies.

Our  model  \cite{12,13}  consists  of  an  $N$-element population
where,  at  each  time  step,  each  element  adopts  one of three
possible  states.  By  analogy  with  epidemiological  SIR  models
\cite{13a},  these  states  are  called  susceptible (S), infected
(I),  and  refractory  (R). The evolution proceeds as follows. At
each  time  step  a  randomly chosen infected element $i$ contacts
another  element  $j$.  Then,  (i)  if  $j$  is in the susceptible
state,  it  becomes  infected;  (ii)  if,  on the contrary, $j$ is
infected  or  refractory,  $i$ becomes refractory. These rules are
better  interpreted  in  the  frame  of a rumor spreading process,
where  S-elements  have  not  heard the rumor yet, I-elements have
heard  the  rumor  and  are willing to transmit it, and R-elements
have  lost  their  interest  in  the rumor and do not transmit it.
Initially,  only  one  element is infected and the remaining $N-1$
elements are susceptible. During the first stage of the evolution,
the  number  of  I-elements  increases.  Since this also implies a
growth  of  the  R-population,  the contacts of I-elements between
themselves  and  with R-elements become more frequent. After a
while, in consequence,  the  I-population  begins to decline.
Eventually, it vanishes   and   the   evolution   stops.   At
the   end,  $N_R$ elements---now  in  the  refractory
state---have been infected at some   stage  during  the
evolution.  Numerical  simulations  and analytical  results
show  that,  generally,  $N_R<N$. Therefore, there  is a fraction
of the population that never hears the rumor. In  the  original
version of this model, each I-element is allowed to  contact  at
random any other element of the population. It has been proven
that, in such situation, the ratio $N_R/N$ approaches a
well-defined  limit,  $N_R/N  = 0.796\dots$, for asymptotically
large values of $N$ \cite{13b,14}.

Here, in contrast, we assume that the elements are situated at the
vertices   of   a   small-world   network,  and  contacts  can  be
established  between linked elements only. The small-world network
is   constructed  from  a  one-dimensional  ordered  network  with
periodic  boundary conditions---a ring---where each node is linked
to  its  $2K$ nearest neighbors, i.e. to the $K$ nearest neighbors
clockwise  and  counterclockwise  \cite{1,2,11}. Then, each of the
$K$  clockwise  connections  of  each  node  $i$  is  rewired with
probability  $p$  to  a randomly chosen node $j$, not belonging to
the  neighborhood  of  $i$.  A  short-cut  between  two  otherwise
distant  regions  is  thus  created. Double and multiple links are
forbidden,  and realizations where the small-world network becomes
disconnected   are  discarded.  As advanced above, the  parameter
$p$ measures the randomness  of  the  resulting  small-world
network. Note that, independently of the value of $p$, the
average number of links per site is always  $2K$. We have
performed series of $10^3$  to $10^5$ numerical realizations of
the model for several values of $p$, $N$, and  $K$.  At  each
realization, the small-world network was generated  anew  and the
evolution was recorded until the exhaustion of the I-population.

\begin{figure}[h]
\resizebox{\columnwidth}{!}{\includegraphics[angle=0]{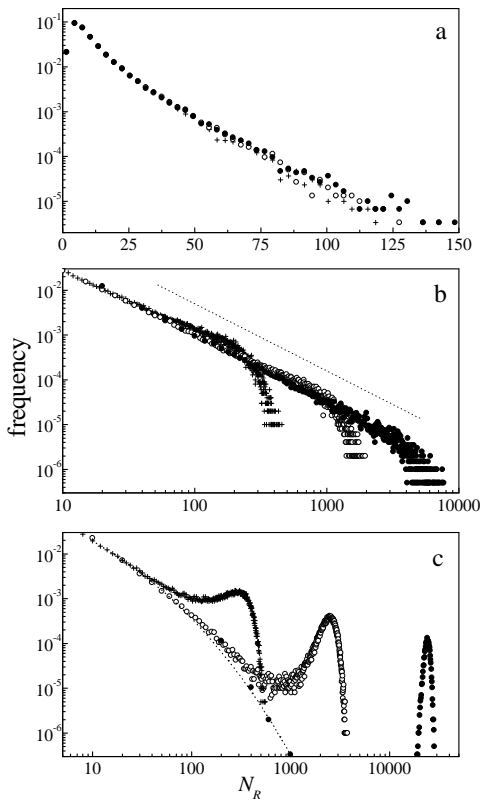}}
\caption{Frequency distribution of the number $N_R$ of R-elements
at the end of the evolution, for $K=2$ and (a) $p=0.05$, (b)
$p=0.19$, and (c) $p=0.3$. Different symbols correspond to
$N=10^3$ (crosses), $N=10^4$ (empty dots), and $N=10^5$ (full
dots). In (b), the dotted straight line has slope   $-1.5$. The
dotted curve in (c) is a schematic representation of the
$N$-independent profile observed for small $N_R$. Note carefully
the different scales of the three plots.}
\label{f1}
\end{figure}

In  the  first  place,  we  have  studied  the distribution of the
number  $N_R$ of R-elements at the end of the evolution for $K=2$.
Figure  \ref{f1} shows the normalized frequency $f(N_R)$ obtained
from series of  $10^5$  realizations  for selected values of $p$
and $N$. For small  randomness  ($p=0.05$,  Fig. \ref{f1}a) the
distribution is approximately  exponential  and  does  not depend
on $N$. In this regime,  the  rumor  ``dies''  after  a few time
steps in a small neighborhood  of  its  origin,  due to the lost
of interest of the highly  interconnected  elements close  to the
initially infected site.   Therefore,  the  size $N$  of  the
whole  population  is irrelevant  to  the  value of $N_R$. The
situation is considerably different   for relatively   large
randomness  ($p=0.3$,  Fig. \ref{f1}c). Here  the  distribution
$f(N_R)$  is bimodal, with a maximum close  to  $N_R  =0$  (not
shown  in  the figure) and an additional  bump for larger $N_R$.
Near $N_R =0$, the frequency is independent   of   $N$,  as  in
the  case  of  small  randomness. Contributions   to   this zone
of  the  distribution  come  from realizations  where
propagation  ceases  before  a  short-cut  is reached.   In
contrast,  the  additional  structure  is  strongly dependent on
$N$.  The  position  of its maximum, in fact, grows linearly,
as   $N_{R}^{\max}\approx   0.25N$.   In   a   typical
realization  contributing  to  this zone of the distribution, many
contacts  occur  through  short-cuts  and  a finite portion of the
population  becomes infected. The intermediate regime, just before
the  large-$N_R$  structure  begins to build up, is illustrated in
Fig. \ref{f1}b  for  $p=0.19$.  The frequency follows here a power
law,  $f(N_R)  \sim N_R^{-\alpha}$ with $\alpha \approx 1.5$, over
a  substantial  interval.  This  interval is limited by above by a
smooth  cut-off,  which  shifts  to  larger  values  of  $N_R$  as
$N^\beta$, with $\beta \sim 0.5$.

\begin{figure}[h]
\resizebox{\columnwidth}{!}{\includegraphics[angle=0]{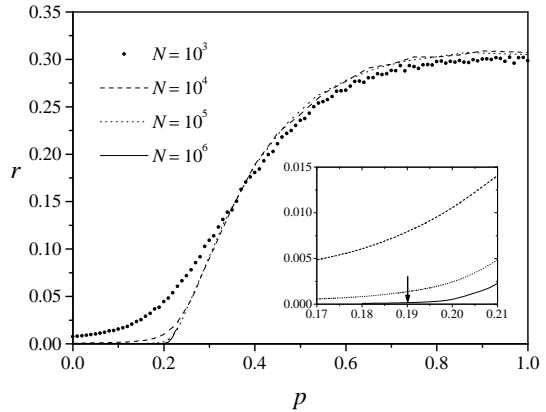}}
\caption{Average fraction $r=\langle N_R/N \rangle$
of refractory elements at the end of the evolution as a function
of the randomness $p$ on a small-world network with $K=2$, and
for $N=10^3$ (dots), $N=10^4$ (dashed line), $N=10^5$ (dotted
line), and $N=10^6$ (full line). The insert shows a close-up in
the transition zone. The arrow indicates the critical value of
$p$, determined as explained in the text.}
\label{f2}
\end{figure}

The  appearance of a well-defined power-law distribution is a clue
to  the  critical-phenomenon  nature  of  the  transition from the
regime  where  the  rumor remains localized to the regime where it
spreads  over a finite fraction of the population. To characterize
the  transition,  we  choose  as  an  order  parameter the average
fraction $N_R/N$  of R-elements at the end of the process,
\begin{equation} \label{r}
r= \langle N_R/N \rangle = N^{-1} \sum_{N_R=0}^N N_R f(N_R),
\end{equation}
which  can  be  straightforwardly  calculated  from  the
numerical results.  In  the  small-randomness  regime, where
$f(N_R)$ is independent of $N$, we expect $r\sim N^{-1}$, so that
$r\to 0$ as $N\to \infty$. For large $p$, in contrast, the
presence of the large-$N_R$ maximum in $f(N_R)$ should provide  a
finite contribution to $r$  even for asymptotically large
populations. These features are verified in numerical
realizations. Figure \ref{f2} shows $r$ as  a function of $p$,
calculated in series of  $10^3$ to $10^4$ realizations for $N$
ranging from $10^3$ to $10^6$ and $K=2$. A well-defined
transition at a finite randomness $p_c \approx 0.2$ is apparent
for large $N$. The  insert  of  Fig. \ref{f2} shows a close-up of
the main plot near  $p_c$.  The smoothness of the curves,
observed even for the largest values of $N$, suggest that the
critical exponent $\gamma$ associated with $r \sim
|p-p_c|^\gamma$ just above the transition, should  be larger than
unity.

\begin{figure}
\resizebox{\columnwidth}{!}{\includegraphics[angle=0]{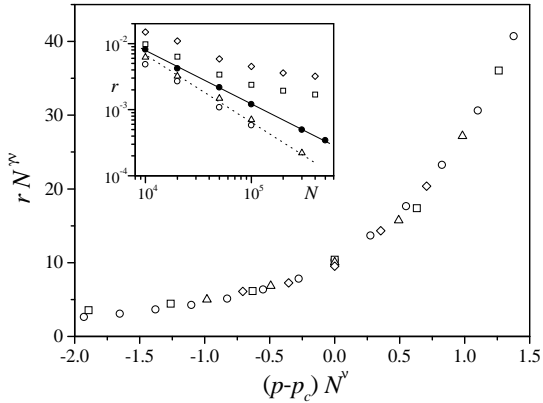}}
\caption{Data collapse in the finite-size scaling of the fraction
$r$ as a function of $p-p_c$, Eq. (\ref{scaling}), with
$\nu=0.36$. Different symbols correspond to different values of
$N$, ranging from $10^4$ to $10^6$. The insert shows $r$ as a
function of $N$ for five values of the randomness: from below,
$p=0.17$,  $0.18$, $0.19$, $0.20$, and $0.21$. The dashed and the
full line have slopes $-1$ and $-0.78$, respectively.}
\label{f3}
\end{figure}

To determine the values of $p_c$ and $\gamma$ we apply finite-size
scaling analysis \cite{fs}. Results are presented in Fig.
\ref{f3}. The insert shows a plot of the fraction $r$ as a
function of $N$ for several values of the randomness $p$. With
$p=0.17$ and $0.18$ we find the subcritical behavior expected for
$p<p_c$, i.e. $r\sim N^{-1}$. For $p=0.20$ and $0.21$, on the
other hand, $r$ saturates to a finite value for very large $N$.
Between these two regimes, with11 $p=0.19$, $r$ decreases as
$r\sim N^{-\rho}$, with $\rho=0.78\pm 0.02$. We identify this
intermediate regime with the critical point, and thus get
$p_c=0.19\pm 0.01$.

Once $p_c$ has been determined, the critical exponent $\gamma$ is
obtained from the finite-size scaling {\it Ansatz} \cite{fs}
\begin{equation} \label{scaling}
rN^{\gamma \nu}=F[(p-p_c)N^{\nu}].
\end{equation}
For $p=p_c$ we have $r=F(0) N^{-\gamma \nu}$, so that $\gamma
\nu= \rho= 0.78\pm 0.02$. The best data collapse near the
critical point is obtained for $\nu=0.36 \pm 0.01$, as shown in
Fig. \ref{f3}. This yields $\gamma=2.2\pm 0.1$.

\begin{figure}
\resizebox{\columnwidth}{!}{\includegraphics[angle=0]{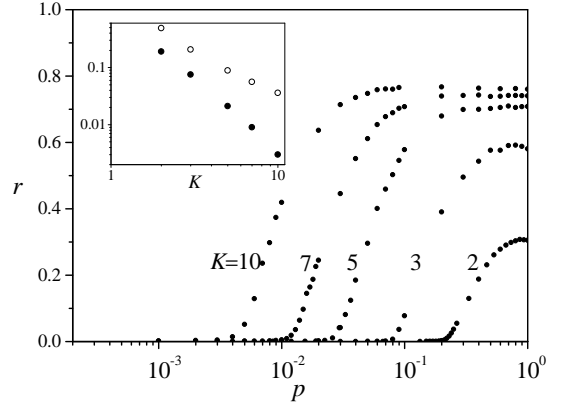}}
\caption{Average fraction $r$ of refractory elements at the end of
the evolution as a function of the randomness $p$ for several
values of $K$, and $N=10^5$.  The insert shows the critical
randomness $p_c$ (full dots) and the  difference $r^*-r_1$ (empty
dots; see text for definitions of $r^*$  and $r_1$) as a function
of $K$.}
\label{f4}
\end{figure}

We have examined this model for other values of $K$, up to $K=10$,
and  found  the  same kind of transition in all cases. The average
fraction  $r$ as a function of $p$ is shown in Figure \ref{f4} for
$N=10^5$  and  several  values  of  $K$,  calculated  over  $10^3$
realizations.  It  is  seen  that  the  critical  randomness $p_c$
decreases  as  $K$  grows.  This  is  due  to the increment in the
number of long-range contacts per element, as a consequence of the
higher  connectivity of each site. On the other hand, the value of
$r$  at  $p  =  1$, $r_1$, grows with $K$ and approaches the level
expected  for  the  original  version  of  the model, $r^* = 0.796
\dots$ \cite{13b,14}.  The  insert  in  Fig. \ref{f4} displays the
dependence of $p_c$  and  of the difference $r^*-r_1$ with $K$.
Though this plot covers  less  that  one  order  of  magnitude in
the $K$-axis, the results  suggest  power-law  decays for both
quantities.

In  summary,  our  numerical analysis shows that the regimes where
the  rumor  is  bounded  to a finite neighborhood of the initially
infected  site  and  where  it  affects  a finite fraction of the
population  are  separated by a well-defined transition at a
finite  randomness $p_c$ of the underlying small-world network.
The results of the finite-size scaling analysis support the
presence of a critical phenomenon at  $p_c$.  Quite
interestingly, the transition is also found if, instead  of
building  a  frozen small-world network as above, a dynamic
small-world  is considered \cite{10}. In this case, each contact
of  an I-element  is established  with  one of its $2K$ nearest
neighbors  with probability  $1-p$,  and with a randomly chosen
site  with probability  $p$. The  transition observed for these
dynamic small-worlds  is  of the  same type as on static
small-world networks,  but  presents quantitative differences.
For $K=2$, for  instance,  the critical  randomness  shifts  to
$p_c \approx 0.07$.

The  geometrical  properties  of  small-world  networks  have been
shown  to  exhibit a cross-over from ordered to random behavior at
$p  \sim  N^{-1}$, which implies a vanishing critical value of $p$
for  asymptotically  large  systems  \cite{5,6,7}. In contrast, we
have  here  presented  evidence that a dynamical process occurring
on  such  structures  exhibits critical behavior at a finite value
of  the  randomness.  This difference suggests that an explanation
for  the  origin  of  the  transition in purely geometrical terms
is unsuitable,  and  specific dynamical properties must be taken
into account.  Very  recently, evidence of critical behavior at
finite small-world randomness  has  been reported for a strictly
epidemiological model, whose  evolution  rules  are substantially
different from those    considered    here \cite{11}.   In
particular,   the epidemiological  model allows  for  the
transformation R $\to$ S, giving  rise to a closed disease cycle
(SIRS) through recovery. Moreover, the transformation  I $\to$ R
$\to$ S is fully deterministic. The critical  phenomenon  found
in that case is a transition to global synchronization   of
local   disease   cycles,  whereas  in  our propagation  process
we  have  a  kind  of percolation phenomenon \cite{8,9}.  In
spite  of these basic differences, the dependence of  the
respective  order  parameters  on  the  randomness $p$ is
strikingly   similar,   even   if  compared  quantitatively.  This
similarity  calls  for  further investigation in order to identify
and   characterize   the  whole  class  of  small-world  dynamical
processes  with  critical  behavior  at  finite randomness, and to
give an analytical description of such behavior.

Enlightening discussions with M. Kuperman and G. Abramson are
gratefully acknowledged.


\begin{thebibliography}{20}
\bibitem{13b}
More precisely, the limit $r^*= \lim_{N\to \infty} N_R/N$ for
random contacts is the solution to the transcendental equation
$r^*=1-\exp(-2r^*)$.
\end{thebibliography}
\end{document}